\documentclass[
aps,
prab,
reprint,
superscriptaddress,
longbibliography,
floatfix
]{revtex4-2}

\usepackage{amsmath,amssymb,amsfonts}
\usepackage{bm}
\usepackage{mathtools}
\usepackage{graphicx}
\usepackage{xcolor}
\usepackage{hyperref}
\usepackage{multirow}

\hypersetup{colorlinks=true,linkcolor=blue,citecolor=blue,urlcolor=blue}

\begin{document}

\title{Superheating field of clean superconductors near the type-I--type-II boundary:\\
the low-temperature Meissner stability limit of niobium}

\author{Takayuki Kubo}
\email[]{kubotaka@post.kek.jp}
\affiliation{High Energy Accelerator Research Organization (KEK), Tsukuba, Ibaraki 305-0801, Japan}
\affiliation{The Graduate University for Advanced Studies (Sokendai), Hayama, Kanagawa 240-0193, Japan}


\begin{abstract}
We calculate the low-temperature superheating field \(B_{\rm sh}\) of clean superconductors near the boundary between type-I and type-II
superconductivity, with particular emphasis on Nb.  
Ginzburg--Landau theory is not applicable at low temperatures, while previous microscopic calculations have been limited to either the large-\(\kappa\) regime or the dirty limit.  
The clean, low-\(\kappa\) regime relevant to Nb has therefore remained unresolved.  
We solve the self-consistent nonlinear nonlocal Eilenberger problem for the Meissner state and then analyze its stability against small
perturbations.
For a Nb-like material with \(\kappa_{\rm GL}=0.7\), 
we obtain \(B_{\rm sh}\simeq290\,{\rm mT}\) at \(T/T_c=0.2\), using \(B_{c0}\simeq200\,{\rm mT}\).  
For a TESLA-shaped Nb accelerator cavity, this corresponds to an intrinsic Meissner-stability limit of about \(67\,{\rm MV/m}\).
\end{abstract}

\maketitle


The superheating field \(B_{\rm sh}\) is the largest magnetic field at which a superconducting surface can remain in the Meissner state~\cite{Fink, Christiansen, Dolgert, Dolgert_erratum, Transtrum, Liarte, Gurevich_2023, Galaiko, Catelani, Lin, Wave, 2021_Kubo,2020_Kubo, 2020_Kubo_erratum}.
For a type-I superconductor, the normal state becomes thermodynamically stable above \(B_c\).  
For a type-II superconductor, vortex states become thermodynamically allowed above \(B_{c1}\).  
In both cases, however, an ideal surface barrier can keep the Meissner state metastable above the
corresponding thermodynamic field. 
The field at which this metastability is lost is \(B_{\rm sh}\), which sets the intrinsic Meissner-stability
limit of a defect-free surface.
This limit is especially important for superconducting radio-frequency (SRF) accelerator cavities~\cite{Padamsee, Gurevich}, 
because these cavities operate in the Meissner state and their intrinsic field limit is expected to be set by \(B_{\rm sh}\). 
For this reason, \(B_{\rm sh}\) of niobium (Nb) has attracted attention in the accelerator community for decades~\cite{JJAP}.

Near the critical temperature \(T_c\), \(B_{\rm sh}\) has been calculated within the Ginzburg--Landau (GL) theory, establishing its dependence on the GL parameter \(\kappa_{\rm GL}\)~\cite{Fink, Christiansen, Dolgert, Dolgert_erratum, Transtrum, Liarte, Gurevich_2023}.
However, GL theory is valid only close to \(T_c\), and therefore cannot determine \(B_{\rm sh}\) at the typical operating temperatures of accelerator cavities, \(T\simeq 2\,{\rm K}\ll T_c\).

To calculate \(B_{\rm sh}\) in this low-temperature regime, one must use a microscopic theory. 
The relevant parameters are the intrinsic value of \(\kappa_0=\lambda_0/\xi_0\), the mean free path \(\ell\), and the temperature \(T\) with \(0<T\le T_c\), 
where \(\lambda_0\) is the clean-limit London penetration depth and \(\xi_0\) is the BCS coherence
length. 
The Eilenberger theory can treat arbitrary \(\ell\) when impurity scattering is included, while the Usadel theory is its diffusive-limit form, applicable for \(\ell\ll\xi_0\).
Several important limits have already been studied (see Table~\ref{tab:summary_hsh}). 
For materials with a large intrinsic value of \(\kappa_0\), such as Nb\(_3\)Sn, 
the Eilenberger problem reduces to a local current response in the large-\(\kappa_0\) limit~\cite{Galaiko, Catelani}. 
Impurity effects in materials with large intrinsic \(\kappa_0\) have also been treated
microscopically~\cite{Lin, Wave}. 
In the dirty limit, the Usadel theory can be used irrespective of the intrinsic value of \(\kappa_0\).
Thus, dirty superconductors with either large or small intrinsic \(\kappa_0\), 
including Nb, have been treated within the Usadel framework \cite{2021_Kubo,2020_Kubo, 2020_Kubo_erratum}.

The remaining problem, which is directly relevant to clean bulk niobium used in SRF cavities, is the low-temperature superheating field of materials with small intrinsic \(\kappa_0\) outside the dirty limit,
namely in the moderately clean or clean regime (see Table~\ref{tab:summary_hsh}).  
In this regime, the penetration depth and the coherence length are comparable, so the current
response is intrinsically nonlocal.  
As a result, nonlinear electrodynamics, self-consistent pair breaking, and Maxwell screening must be treated on the same footing. 
This requires solving the spatially resolved nonlinear nonlocal Eilenberger problem. 
This Letter addresses the clean-limit part of this problem by solving the spatially resolved nonlinear nonlocal Eilenberger stability problem. 
We focus on the Nb-relevant range \(0.7\lesssim\kappa_{\rm GL}\lesssim1\), or equivalently
\(0.73\lesssim\kappa_0\lesssim1.04\), 
which is representative of clean niobium-like materials relevant to SRF accelerator cavities.

Note that \(\kappa_0\) and \(\kappa_{\rm GL}\) are different quantities. 
The GL parameter \(\kappa_{\rm GL}\) is defined in the vicinity of \(T_c\), 
although its value is temperature independent within GL theory.
In the clean limit, it can be expressed in terms of the microscopic parameter \(\kappa_0=\lambda_0/\xi_0\) through \(\kappa_{\rm GL}=\kappa_0/\sqrt{2K_c}\), with \(K_c=7\zeta(3)\pi^2/(48e^{2\gamma_E})\). Here \(\gamma_E\simeq0.577\) is the Euler constant.
Numerically, \(\kappa_{\rm GL}\simeq0.958\kappa_0\).


\begin{table}[t]
\caption{
Summary of microscopic theories for the superheating field in the low-temperature regime relevant to SRF cavities. 
Here \(\lambda_0\) and \(\xi_0\) are the clean-limit London penetration depth and BCS
coherence length, respectively, and define the intrinsic material parameter \(\kappa_0=\lambda_0/\xi_0\). 
The mean free path \(\ell\) is an extrinsic parameter controlled by nonmagnetic impurity scattering,
with shorter \(\ell\) corresponding to stronger scattering. 
}
\label{tab:summary_hsh}
\begin{ruledtabular}
\renewcommand{\arraystretch}{1.25}
\setlength{\tabcolsep}{3pt}
\begin{tabular}{
p{0.29\columnwidth}
p{0.31\columnwidth}
p{0.33\columnwidth}
}
Intrinsic
\(\lambda_0/\xi_0\)
&
Mean free path $\ell$

&
Theory
\\
\hline
\multirow{3}{0.25\columnwidth}{
\(\lambda_0/\xi_0\gg1\)
\newline
(e.g., Nb\(_3\)Sn, NbN)
}
&
Clean limit
\newline
\((\ell\gg\xi_0)\)
&
Refs.~\cite{Galaiko, Catelani, Lin}.
\(B_{\rm sh}(0)/B_{c0}=0.84\).
\\
\cline{2-3}
&
Intermediate 
\newline
\((\ell \sim \xi_0)\)
&
Ref.~\cite{Lin}.
\( 0.84 \lesssim B_{\rm sh}(0)/B_{c0}\lesssim 0.88\).
\\
\cline{2-3}
&
\multirow{2}{0.23\columnwidth}{
Dirty limit
\newline
\((\ell\ll\xi_0)\)
}
&
\multirow{2}{0.33\columnwidth}{
Refs.~\cite{Lin, 2021_Kubo, 2020_Kubo}.
\(B_{\rm sh}(0)/B_{c0}\simeq 0.8\).
}
\\
\cline{1-1}
\multirow{3}{0.25\columnwidth}{
\(\lambda_0/\xi_0\lesssim1\)
\newline
(e.g., Nb)
}
&
&
\\
\cline{2-3}
&
Intermediate
\newline
\((\ell \sim \xi_0)\)
&
Not available
\\
\cline{2-3}
&
Clean limit
\newline
\((\ell\gg\xi_0)\)
&
{\bf Present work}.
\end{tabular}
\end{ruledtabular}
\end{table}

\begin{figure}[t]
\includegraphics[width=0.8\linewidth]{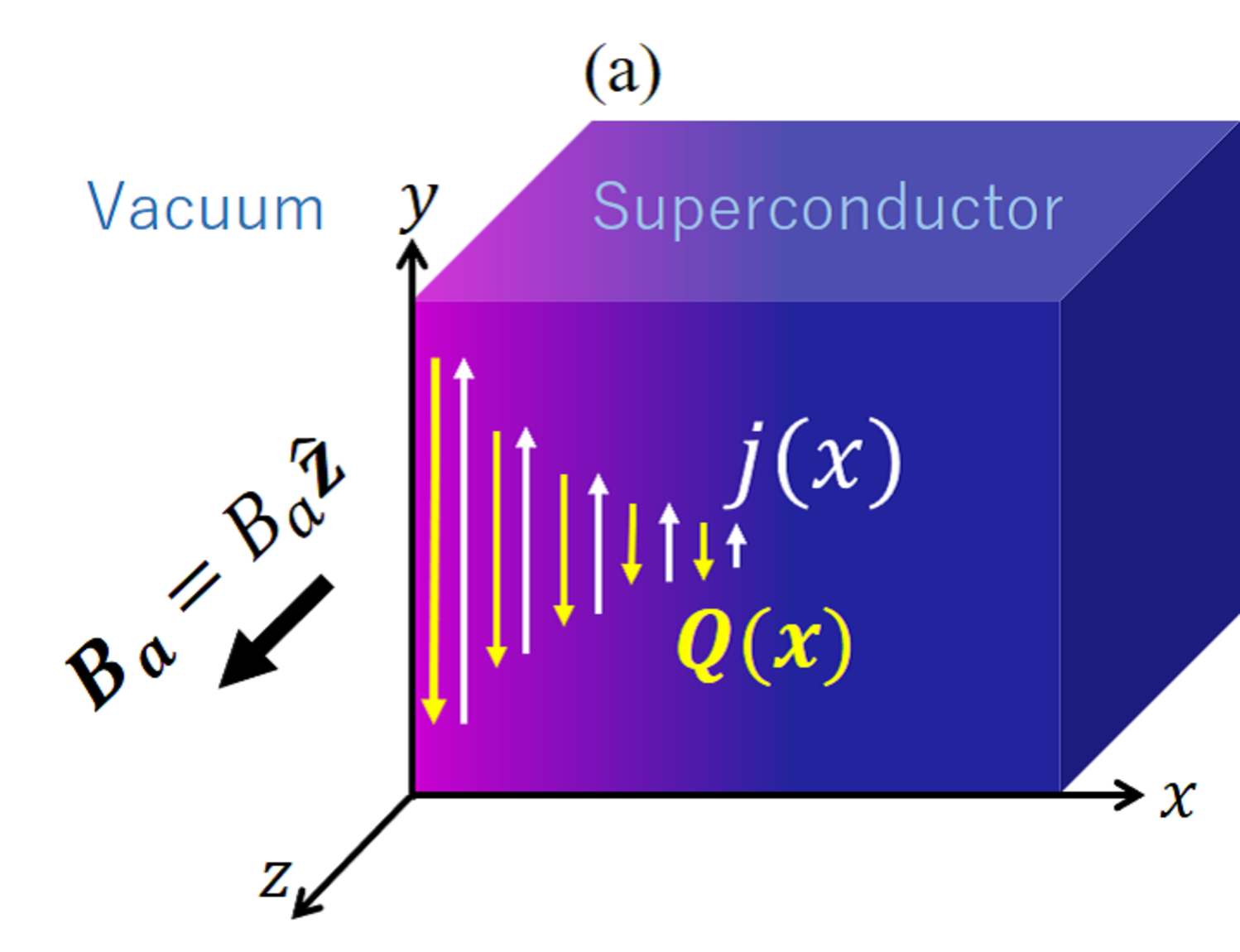}
\includegraphics[width=0.8\linewidth]{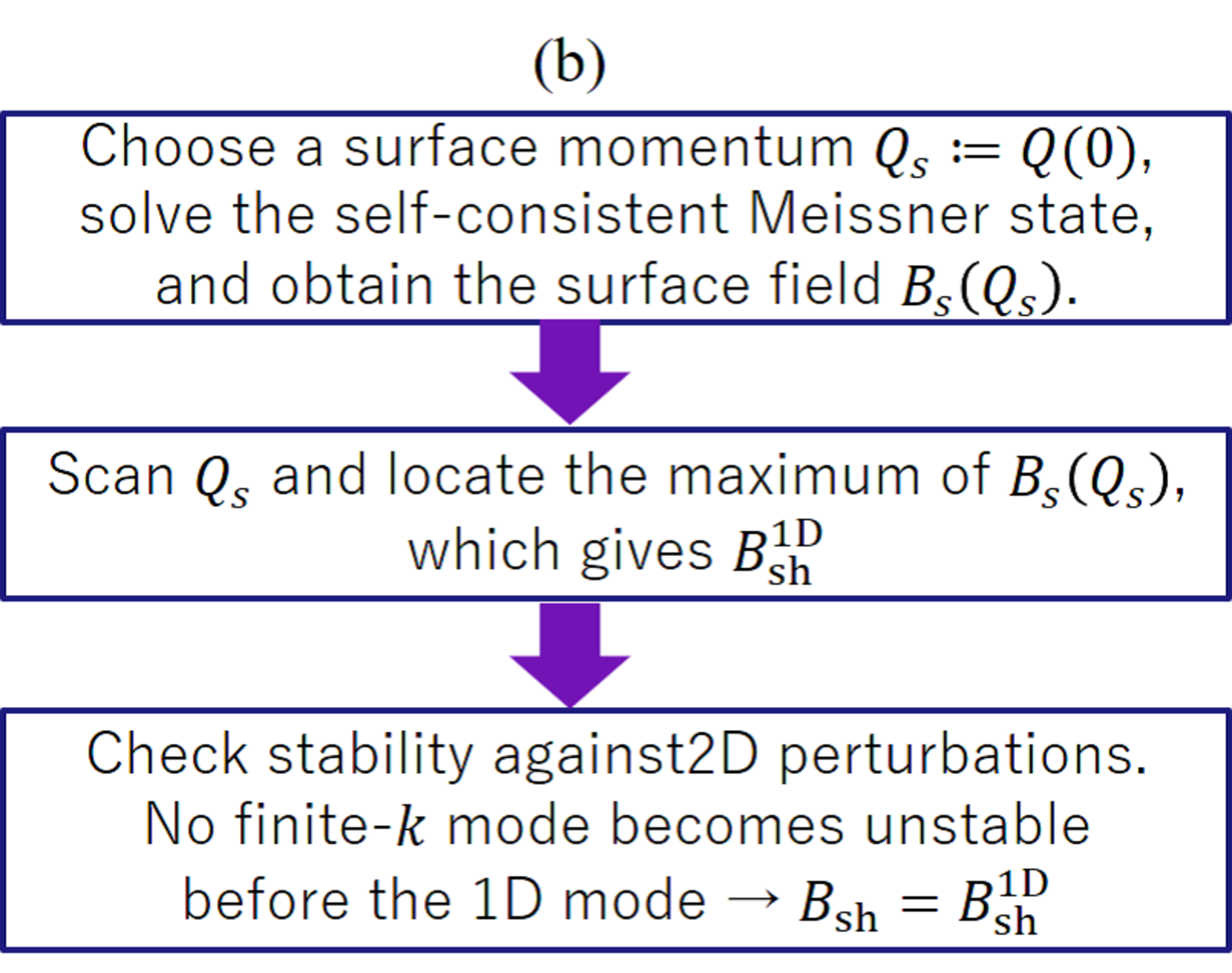}
\caption{
(a) Geometry of the Meissner state in a semi-infinite superconductor.
The applied magnetic field is along \(z\), while the screening current and superfluid momentum are parallel to the
\(y\) axis.
(b) Procedure for determining \(B_{\rm sh}\).  See the Supplemental Material~\cite{SM} for details.
}
\label{fig1}
\end{figure}

We consider a clean semi-infinite superconductor occupying \(x>0\), as shown in Fig.~\ref{fig1}(a).
The applied dc magnetic field is parallel to the surface,
\({\bf B}_a=B_a\hat{\bf z}\), and the Meissner screening current flows along \(\hat{\bf y}\).  
Energies are normalized by the zero-temperature BCS gap \(\Delta_0\), and lengths by \(\xi_0=\hbar v_F/\pi\Delta_0\). 
The magnetic flux density is normalized by \(B_{c0}=\Delta_0\sqrt{\mu_0N_0}\), and the current density
by \(J_0=H_{c0}/\lambda_0\), where \(B_{c0}=\mu_0H_{c0}\) and \(\lambda_0^{-2}=(2/3)\mu_0e^2N_0v_F^2\). 
Here \(N_0\) is the single-spin normal-state density of states at the Fermi level.
We define \(\kappa_0=\lambda_0/\xi_0\), the dimensionless temperature \(t=k_BT/\Delta_0\), the dimensionless critical temperature \(t_{c0}=k_BT_c/\Delta_0=e^{\gamma_E}/\pi\), and the dimensionless
Matsubara frequencies \(\Omega_n=\hbar\omega_n/\Delta_0=(2n+1)\pi t\). 
We use the dimensionless coordinates \(\bar x=x/\xi_0\) and \(\bar y=y/\xi_0\).  
The dimensionless gauge-invariant wave number is ${\bf Q}=\xi_0 \{ \nabla\chi+(2\pi/\phi_0){\bf A} \}$.

The nonlinear nonlocal Meissner problem is defined by
\begin{eqnarray}
&&\Bigl[ \Omega_n+\frac{\pi}{2}\hat{\bf v}\cdot (\bar\nabla+i{\bf Q}) \Bigr] f_n
= \Delta g_n, \label{eq:eilenberger_f} \\
&&\Bigl[ \Omega_n-\frac{\pi}{2}\hat{\bf v}\cdot (\bar\nabla-i{\bf Q}) \Bigr] f_n^\dagger
= \Delta^* g_n, \label{eq:eilenberger_fd} \\
&& g_n^2+f_nf_n^\dagger = 1, \label{eq:eilenberger_norm} \\
&& \Delta\ln\frac{t}{t_{c0}} + 2\pi t\sum_{n\ge0} \Bigl( \frac{\Delta}{\Omega_n}
- \langle f_n\rangle \Bigr) = 0, \label{eq:gap} \\
&&{\bf j} =-2\pi\sqrt{6}\,t \sum_{n\ge0} \langle \hat{\bf v}\,{\rm Im}\,g_n \rangle,
\label{eq:current} \\
&&\bar\nabla\times{\bf Q} = \frac{\sqrt{6}}{\pi\kappa_0}{\bf b}, \label{eq:maxwell_Q_b} \\
&&\bar\nabla\times{\bf b} = \frac{1}{\kappa_0}{\bf j}. \label{eq:maxwell_b_j}
\end{eqnarray}
Here \(\bar\nabla=\xi_0\nabla\), and \(\langle\cdots\rangle\) denotes the Fermi-surface average.
Equations~(\ref{eq:eilenberger_f}) and (\ref{eq:eilenberger_fd}) are the clean Eilenberger equations,
Eq.~(\ref{eq:eilenberger_norm}) is the normalization condition, 
Eq.~(\ref{eq:gap}) is the gap equation,
Eq.~(\ref{eq:current}) gives the nonlinear current response, and
Eqs.~(\ref{eq:maxwell_Q_b}) and (\ref{eq:maxwell_b_j}) are Maxwell's equations. 
For the one-dimensional Meissner state, \(\Delta=\Delta(\bar x)\),
\({\bf Q}=Q(\bar x)\hat{\bf y}\), and
\({\bf b}=b(\bar x)\hat{\bf z}\), 
we impose \(Q'(0)=(\sqrt{6}/\pi\kappa_0)b_a\), \(Q(\infty)=0\), and \(\Delta(\infty)=\Delta_b(T)\), 
where \(\Delta_b(T)\) is the zero-field bulk gap. 
The quasiclassical propagators satisfy specular reflection at the surface,
\({\cal F}_n(\hat v_x,\hat v_y,\hat v_z;0)
={\cal F}_n(-\hat v_x,\hat v_y,\hat v_z;0)\), with
\({\cal F}_n=f_n,f_n^\dagger,g_n\).

It is useful to use the Riccati parametrization
\(f_n=2a_n/(1+a_nb_n)\), \(f_n^\dagger=2b_n/(1+a_nb_n)\), and \(g_n=(1-a_nb_n)/(1+a_nb_n)\).  
For the one-dimensional Meissner solution, 
\(\Delta=\Delta(\bar x)\) and \({\bf Q}=Q(\bar x)\hat{\bf y}\), 
the Eilenberger equations reduce to 
$\pi\hat v_x a_n' = \Delta(1-a_n^2)- (2\Omega_n+i\pi\hat v_yQ )a_n$ and $\pi\hat v_x b_n' = -\Delta(1-b_n^2) + (2\Omega_n+i\pi\hat v_yQ )b_n$. 
Here the prime denotes differentiation with respect to \(\bar x\).


To determine \(B_{\rm sh}\), we proceed in three steps.  
The geometry and the overall calculation procedure are summarized in Fig.~\ref{fig1}.  
We first construct self-consistent nonlinear Meissner-state solutions.  
It is convenient to specify the
surface superfluid momentum \(Q_s=Q(0)\).  For each value of \(Q_s\),
Eqs.~(\ref{eq:eilenberger_f})--(\ref{eq:maxwell_b_j}) are solved
self-consistently to obtain the spatial profiles of the pair potential,
superfluid momentum, screening current, and magnetic field.  The
corresponding surface magnetic field \(B_s(Q_s)\) is then obtained from
the derivative of \(Q\) at the surface.  The detailed construction of
these nonlinear Meissner-state solutions is described in Sec.~I of the
Supplemental Material~\cite{SM}.

We next examine the stability of each Meissner-state solution against
small perturbations.  The superheating field is the first applied field
at which the Meissner state loses stability.  We consider a perturbation
with wave number $k$ along the surface,
\begin{eqnarray}
\delta\Delta(\bar x,\bar y)
&=&
\eta(\bar x)\cos \bar k\bar y,
\label{eq:dDelta}
\\
\delta{\bf Q}(\bar x,\bar y)
&=&
W(\bar x)\sin \bar k\bar y\,\hat{\bf x}
+
V(\bar x)\cos \bar k\bar y\,\hat{\bf y}.
\label{eq:dQ}
\end{eqnarray}
Here, \(\bar k=k\xi_0\) is the dimensionless wave number.
The corresponding magnetic-field perturbation follows from
\(\bar\nabla\times\delta{\bf Q}
=(\sqrt{6}/\pi\kappa_0)\delta{\bf b}\).
We write
\(\delta b_z(\bar x,\bar y)
=(\pi\kappa_0/\sqrt{6})h(\bar x)\cos\bar k\bar y\),
where \(h=V'-\bar kW\), and the prime denotes differentiation with
respect to \(\bar x\).  At fixed applied field, the magnetic
perturbation must satisfy \(h(0)=0\), and all perturbations vanish as
\(\bar x\to\infty\).

We first determine the one-dimensional stability limit
\(B_{\rm sh}^{\rm 1D}\), corresponding to \(\bar k=0\) in
Eqs.~(\ref{eq:dDelta}) and (\ref{eq:dQ}).  This choice is motivated by
the well-known GL result that the superheating field is determined by
the one-dimensional instability for
\(\kappa_{\rm GL}\lesssim1.1\).  For the one-dimensional case, the
stability problem is significantly simplified.  As shown in Sec.~II of
the Supplemental Material~\cite{SM}, the first maximum of \(B_s(Q_s)\) is the
point at which the Meissner state loses stability.  Therefore the
one-dimensional stability limit \(B_{\rm sh}^{\rm 1D}\) can be obtained
directly by scanning \(Q_s\) and locating the first maximum of
\(B_s(Q_s)\), corresponding to
\(B_{\rm sh}^{\rm 1D}=\max_{Q_s}B_s(Q_s)\)
[Eq.~(S39) of the Supplemental Material].

The first unstable perturbation, however, does not necessarily have to
be one-dimensional.  Although the Meissner-state solution is uniform
along the surface, an unstable perturbation may vary along the surface.
We therefore also examine perturbations with finite \(\bar k\).
For a given Meissner-state solution, we linearize the Riccati, gap, and
Maxwell equations with respect to the perturbations in
Eqs.~(\ref{eq:dDelta}) and (\ref{eq:dQ}).  After discretization on the
\(\bar x\)-mesh, the perturbation amplitudes form the vector
\({\bf u}=(\eta_1,\ldots,\eta_N,W_1,\ldots,W_N,
V_1,\ldots,V_N)^T\).  The linearized Riccati equations determine the
induced Green functions, from which we construct the residuals of the
linearized gap and Maxwell equations.  This defines the discretized
stability operator \({\cal M}_{\bar k}\).  An instability occurs when a
nonzero perturbation satisfies all the linearized equations and
boundary conditions, namely when
\({\cal M}_{\bar k}{\bf u}=0\) has a nontrivial solution.  Numerically,
this condition is detected when the smallest singular value of
\({\cal M}_{\bar k}\) approaches zero.  The finite-\(\bar k\) stability
analysis is described in Sec.~III of the Supplemental Material~\cite{SM}.

In the parameter space \((\kappa_0,T)\) studied here, we find no
finite-\(\bar k\) instability below \(B_{\rm sh}^{\rm 1D}\).  Hence we
find that, throughout the parameter range studied here, the
superheating field is given by the one-dimensional value,
\(B_{\rm sh}=B_{\rm sh}^{\rm 1D}\).


\begin{figure}[t]
\includegraphics[width=\linewidth]{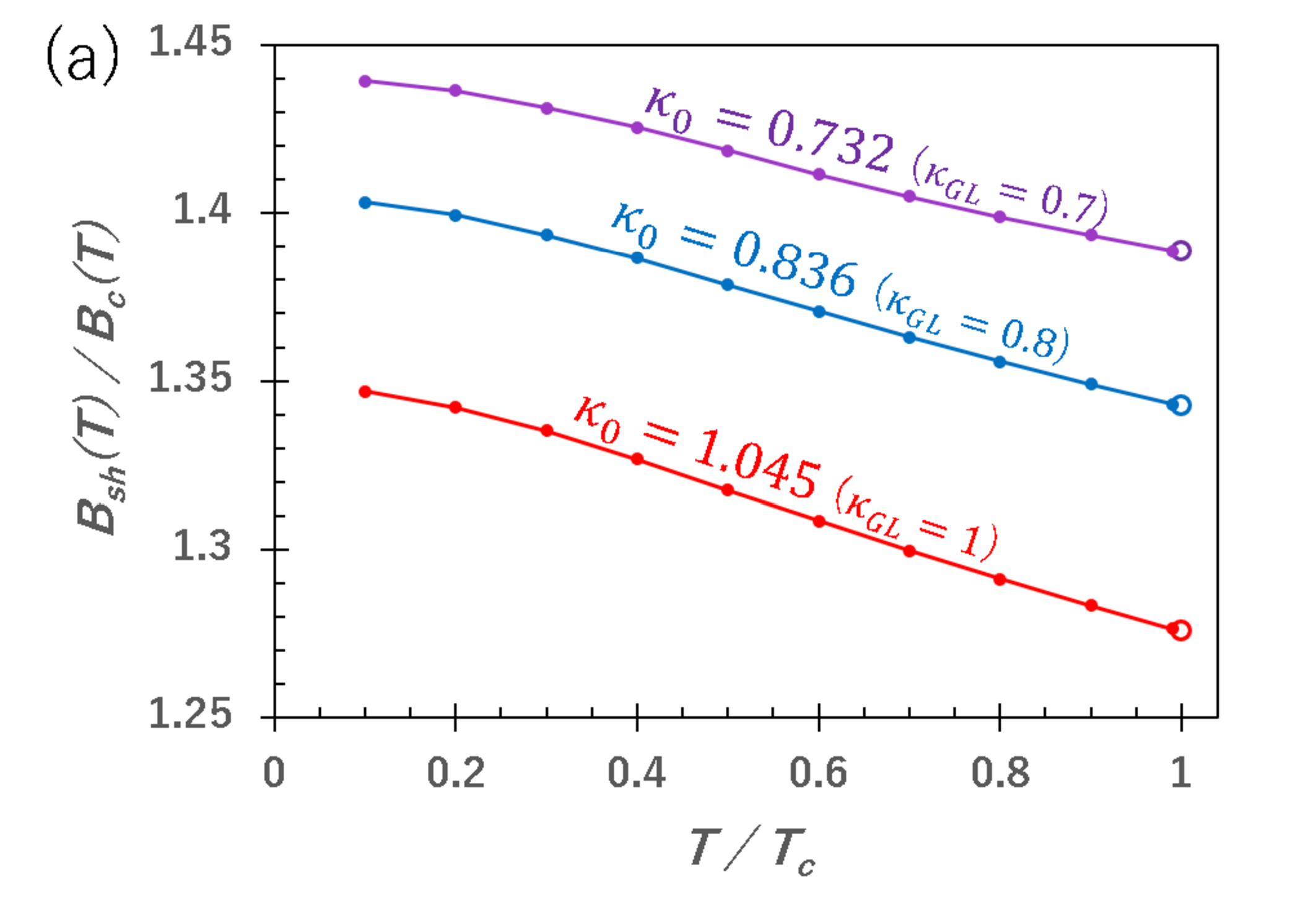}
\includegraphics[width=\linewidth]{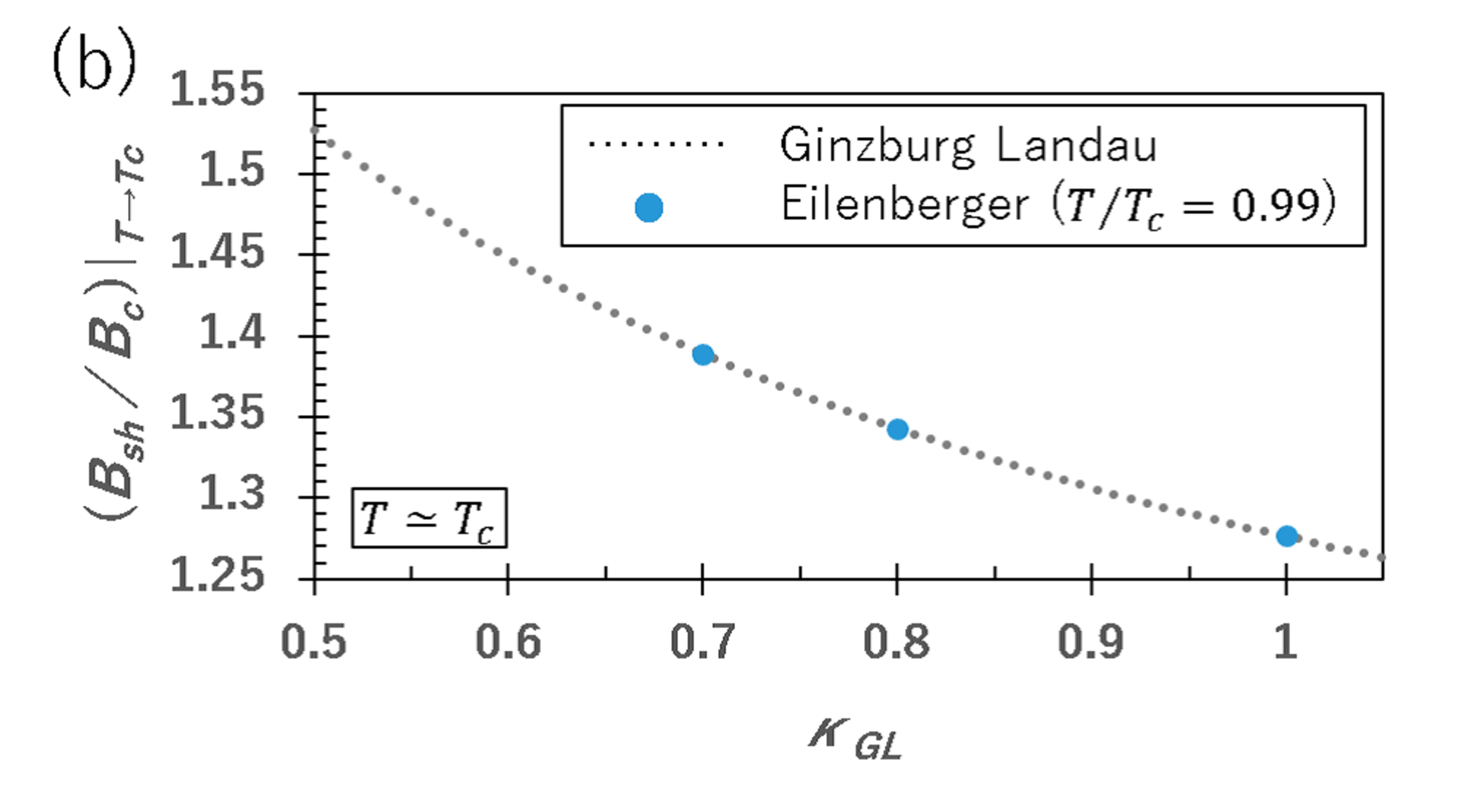}
\caption{
(a) Temperature dependence of the superheating field for a clean low-\(\kappa\) superconductor. 
Filled symbols show Eilenberger results at fixed \(\kappa_0\); the
corresponding \(\kappa_{\rm GL}\) values are obtained from the near-\(T_c\)
mapping. 
Open symbols at \(T/T_c=1\) show the corresponding GL values, which are valid in the limit \(T\to T_c\). 
(b) Superheating field near \(T_c\) as a function of \(\kappa_{\rm GL}\). 
The agreement with the GL result provides a check of the Eilenberger calculation in the near-\(T_c\) limit.
}
\label{fig2}
\end{figure}

Figure~\ref{fig2} (a) shows the temperature dependence of \(B_{\rm sh}(T)/B_c(T)\) for several fixed values of \(\kappa_0=\lambda_0/\xi_0\), calculated from the self-consistent nonlocal Eilenberger theory. 
This microscopic formulation is applicable over the whole temperature range \(0<T\le T_c\). 
The results at \(T/T_c=0.99\) agree well with the corresponding Ginzburg--Landau values shown by the open symbols at \(T/T_c=1\), confirming that the Eilenberger calculation correctly reproduces the near-\(T_c\) GL limit [see also Figure~\ref{fig2} (b)]. 
At lower temperatures, the GL expansion is no longer valid, and the nonlocal nonlinear Eilenberger problem must be solved.  
In this regime, \(B_{\rm sh}/B_c\) becomes larger than one would expect
from a simple extrapolation of the GL result near \(T_c\).  A similar
low-temperature enhancement is found in previous microscopic
calculations in both the clean and dirty large-\(\kappa\) limits.  In
the clean large-\(\kappa\) limit, \(B_{\rm sh}/B_c\) increases from
about \(0.745\) near \(T_c\) to about \(0.84\) at \(T=0\)~\cite{Galaiko,Catelani,Lin,Wave}, corresponding
to an enhancement by a factor of about \(1.12\).  In the dirty limit,
the corresponding value at \(T=0\) is about \(0.8\)~\cite{Lin,2021_Kubo}, giving an
enhancement by a factor of about \(1.07\).  In the present
low-\(\kappa\) regime, the enhancement is of similar magnitude: for
\(\kappa_{\rm GL}\simeq1\), \(B_{\rm sh}/B_c\) increases from about
\(1.27\) near \(T_c\) to about \(1.35\) at low temperature, corresponding
to a factor of about \(1.06\).


We now estimate the intrinsic field limit of clean Nb.  
Multiple experiments on high-purity Nb support its type-II/1 character.
Small-angle neutron scattering measurements on Nb with \( {\rm RRR}>10^4\) observed a well-ordered vortex lattice that transforms on cooling into an intermediate mixed state composed of dense vortex-lattice domains and Meissner regions~\cite{Backs}. 
Magneto-optical imaging of high-purity cavity-grade Nb (\( {\rm RRR}\simeq 500\)) also observed real-space phase separation into vortex bundles and Meissner regions, and single-vortex-resolved imaging has directly shown clustering of attractively interacting vortices~\cite{Ooi_1,Ooi_2}. 
These results, together with recent low-energy muon spin spectroscopy/secondary-ion mass spectrometry measurements giving \(\kappa_{\rm GL}\simeq0.7\), support the view that clean Nb lies very close to the type-I--type-II boundary \cite{Junginger}.
For this value, \(\kappa_{\rm GL}=0.7\), we find \(B_{\rm sh}(T/T_c=0.2)\simeq1.44B_{c0}\).  
Using \(B_c(T/T_c=0.2)\simeq B_{c0}\simeq200\,{\rm mT}\), this gives \(B_{\rm sh}^{\rm Nb}\simeq290\,{\rm mT}\). 
For accelerator cavities, this surface magnetic-field limit can be converted into a theoretical limit on the accelerating gradient. 
Using the TESLA-shape conversion factor \(4.26\,{\rm mT}/({\rm MV/m})\), 
we obtain \(E_{\rm acc}^{\rm max}\simeq 67\,{\rm MV/m}\). 
The predicted limit, \(B_{\rm sh}\simeq290\,{\rm mT}\), corresponding
to \(E_{\rm acc}^{\rm max}\simeq67\,{\rm MV/m}\), is substantially
higher than the best performance achieved in Nb cavities, with peak
surface magnetic fields of about \(200\,{\rm mT}\), corresponding to
accelerating gradients of about \(45\)--\(50\,{\rm MV/m}\).
A summary of the experimental data is provided in Ref.~\cite{JJAP};
see also the references therein.  
This discrepancy may arise from various extrinsic limitations in real cavities,
including suppression of \(B_{\rm sh}\) by current crowding at rough surfaces
(see, e.g., Refs.~\cite{Kubo_2015,Lechner} and references therein)
and local magnetic-field enhancement at defects
(see, e.g., Refs.~\cite{Iwashita,Kubo_pit,Xie} and references therein),
which are not included in the present ideal-surface calculation.

At lower temperatures, the numerical cost increases rapidly because the Matsubara spacing becomes smaller and the angular integrands become sharper. 
The latter is caused by the weak Matsubara smoothing of the Doppler-shifted quasiparticle response in the strongly current-carrying Meissner state. 
The present calculation was therefore limited to \(T/T_c\ge 0.1\).


We have formulated and solved, for the first time, the superheating-field problem for clean low-\(\kappa\) superconductors as a nonlinear nonlocal Eilenberger stability problem.  
The formulation is valid over \(0<T\le T_c\), and the present numerical results cover \(T/T_c\ge0.1\).
This regime had not been accessible from the previously available limits, as summarized in Table~\ref{tab:summary_hsh}.
For SRF accelerator cavities, the result provides a microscopic estimate of the intrinsic field limit of clean niobium-like materials. 
This limit serves as a reference value for separating the ideal Meissner-stability limit from practical limitations caused by defects, heating, trapped flux, and other extrinsic effects.

The present formulation also serves as a starting point for future microscopic studies of SRF multilayers on clean Nb substrates. 
Several theoretical models have been developed for multilayer coatings~\cite{Kubo_ML1,Kubo_ML2,2021_Kubo,Gurevich_ML1,Gurevich_ML2,Posen_ML,Gurevich_2023}.
However, a low-temperature microscopic calculation has not yet been developed for a multilayer system that explicitly includes the nonlocal Eilenberger response of a clean Nb substrate. 
The present work determines the nonlocal response and Meissner-stability limit of the clean Nb substrate itself, which can be coupled to superconducting coating layers in future multilayer calculations.

\begin{acknowledgments}
This work was supported by the U.S.-Japan Science and Technology Cooperation Program in High Energy Physics under Grant No. 2026-15-1, and by JSPS KAKENHI under Grant Nos. JP26K03209 and JP26K00665.
\end{acknowledgments}

\clearpage
\onecolumngrid
\setcounter{section}{0}
\setcounter{subsection}{0}
\setcounter{equation}{0}
\renewcommand{\theequation}{S\arabic{equation}}
\renewcommand{\theHequation}{supp.\arabic{equation}}
\renewcommand{\theHsection}{supp.\arabic{section}}
\renewcommand{\theHsubsection}{supp.\arabic{section}.\arabic{subsection}}
\begin{center}
{\large\bfseries Supplemental Material for ``Superheating field of clean superconductors near the type-I--type-II boundary: the low-temperature Meissner stability limit of niobium''\par}
\vspace{9pt}
Takayuki Kubo\\[4pt]
{\small High Energy Accelerator Research Organization (KEK), Tsukuba, Ibaraki 305-0801, Japan}\\
{\small The Graduate University for Advanced Studies (Sokendai), Hayama, Kanagawa 240-0193, Japan}
\end{center}
\vspace{1em}
\twocolumngrid

\section{Nonlinear nonlocal Meissner solution}
\label{sec:nonlinear_meissner}

Before considering the stability problem, we first construct
self-consistent Meissner-state solutions for specified values of the
surface superfluid momentum,
\begin{eqnarray}
Q_s:=Q(0).
\end{eqnarray}
For each value of \(Q_s\), the calculation determines the spatial
profiles of the order parameter, superfluid momentum, current density,
and magnetic field.  After the self-consistent solution is obtained, the
corresponding surface magnetic field \(b_a:=b(0)\) is calculated from
\begin{eqnarray}
b_a=\frac{\pi\kappa_0}{\sqrt{6}}Q'(0).
\label{eq:S_ba_Qs}
\end{eqnarray}
We identify the surface value of the magnetic field, \(b(0)\), with the
applied field \(b_a\) just outside the superconductor.  Thus, by varying
\(Q_s\), we obtain the relation between the surface superfluid momentum
and the surface magnetic field, \(b_a(Q_s)\).  The stability analysis is
performed only after these Meissner-state solutions have been obtained.

We use the dimensionless variables defined in the Letter:
\(\bar x=x/\xi_0\), \(t=k_BT/\Delta_0\),
\(\Omega_n=(2n+1)\pi t\), and \(b=B/B_{c0}\).  Here
\(t_{c0}=k_BT_{c0}/\Delta_0=e^{\gamma_E}/\pi\), and
\(\Delta_{\rm b}(t)\) denotes the zero-current bulk gap at temperature
\(t\).  The clean Eilenberger equations are solved for a semi-infinite
superconductor occupying
\(\bar x>0\).  The surface is at \(\bar x=0\), the applied magnetic
field is along \(\hat{\bf z}\), and the screening current flows along
\(\hat{\bf y}\).  The Meissner-state solution depends only on \(\bar x\)
and is written as
\begin{eqnarray}
\Delta=\Delta(\bar x),
\qquad
{\bf Q}=Q(\bar x)\hat{\bf y},
\qquad
{\bf b}=b(\bar x)\hat{\bf z}.
\end{eqnarray}
Here \(\Delta\) is chosen real, and the phase gradient is included in
\({\bf Q}\).  With the Riccati parametrization of the Matsubara Green's
functions,
\begin{eqnarray}
f_n&=&\frac{2a_n}{1+a_nb_n},
\qquad
f_n^\dagger=\frac{2b_n}{1+a_nb_n}, \\
g_n&=&\frac{1-a_nb_n}{1+a_nb_n},
\end{eqnarray}
the Eilenberger equations become
\begin{eqnarray}
\pi\hat v_x a_n'
&=&
\Delta(1-a_n^2)-\left(2\Omega_n+i\pi\hat v_yQ\right)a_n,
\label{eq:S_a_meissner}\\
\pi\hat v_x b_n'
&=&
-\Delta(1-b_n^2)+\left(2\Omega_n+i\pi\hat v_yQ\right)b_n .
\label{eq:S_b_meissner}
\end{eqnarray}
The prime denotes differentiation with respect to \(\bar x\).  The
amplitudes \(a_n\) and \(b_n\) also depend on the Fermi-velocity
direction \(\hat{\bf v}\) and on the Matsubara index \(n\), although
these arguments are not written explicitly.
Here \(\hat{\bf v}={\bf v}_F/v_F\) denotes the direction of the Fermi
velocity, and \(\hat v_x\), \(\hat v_y\), and \(\hat v_z\) are its scalar
components.
At the far end of the computational interval, \(\bar x=\bar L\), the
solution is matched to the zero-current bulk state:
\begin{eqnarray}
\Delta(\bar L)=\Delta_{\rm b}(t),
\qquad
Q(\bar L)=0.
\label{eq:S_bulk_fields}
\end{eqnarray}
The corresponding bulk Riccati amplitudes are
\begin{eqnarray}
a_{n,{\rm b}}=b_{n,{\rm b}}
=
\frac{\Delta_{\rm b}(t)}
{\Omega_n+\sqrt{\Omega_n^2+\Delta_{\rm b}^2(t)}} .
\label{eq:S_bulk_Riccati}
\end{eqnarray}
Although the Meissner-state solution depends only on \(\bar x\), the
Riccati amplitudes still depend on the Fermi-velocity direction
\(\hat{\bf v}\).  Equivalently, one may parametrize the Fermi-velocity
direction by two angles on the Fermi surface and write
\(a_n=a_n(\bar x;\theta,\phi)\) and
\(b_n=b_n(\bar x;\theta,\phi)\).  The scalar component \(\hat v_x\)
enters through the derivative terms \(\pi\hat v_x a_n'\) and
\(\pi\hat v_x b_n'\), while the scalar component \(\hat v_y\) enters
through the Doppler-shift term proportional to \(\hat v_yQ\).

For positive Matsubara frequencies, the stable integration direction is
chosen according to the sign of \(\hat v_x\).  For \(\hat v_x>0\),
\(a_n\) is propagated from the surface toward the bulk, whereas \(b_n\)
is propagated from the bulk toward the surface.  For \(\hat v_x<0\), the
directions are reversed: \(a_n\) is propagated from the bulk toward the
surface, whereas \(b_n\) is propagated from the surface toward the bulk.
Integrating in the opposite direction would make small numerical errors
grow exponentially.
The calculation is organized by pairing directions with opposite normal
components, \(\hat v_x\) and \(-\hat v_x\).  The amplitudes whose stable
integration starts from the bulk are obtained first, using the bulk
values in Eq.~(\ref{eq:S_bulk_Riccati}).  For positive Matsubara
frequencies, these are \(a_n\) for \(\hat v_x<0\) and \(b_n\) for
\(\hat v_x>0\).  The remaining amplitudes start from the surface.  Their
initial values are supplied by the specular-reflection condition at
\(\bar x=0\), as described below.

At the surface, \(\bar x=0\), we impose specular reflection.  The
reflected direction has the same tangential velocity components and the
opposite normal velocity component.  Therefore,
\begin{eqnarray}
{\cal F}_n(0;\hat v_x,\hat v_y,\hat v_z)
=
{\cal F}_n(0;-\hat v_x,\hat v_y,\hat v_z),
\label{eq:S_specular_Green}
\end{eqnarray}
where \({\cal F}_n\) denotes \(f_n\), \(f_n^\dagger\), or \(g_n\).
Equivalently, in the Riccati representation, the incoming and reflected
amplitudes are connected by
\begin{eqnarray}
a_n(0;\hat v_x,\hat v_y,\hat v_z)
&=&
a_n(0;-\hat v_x,\hat v_y,\hat v_z),
\label{eq:S_specular_a}\\
b_n(0;\hat v_x,\hat v_y,\hat v_z)
&=&
b_n(0;-\hat v_x,\hat v_y,\hat v_z).
\label{eq:S_specular_b}
\end{eqnarray}
No independent boundary condition is imposed on \(\Delta\) at the
surface.  Its surface value is determined self-consistently by the gap
equation together with the reflected quasiparticle trajectories.  In the
GL limit, this prescription reduces to the usual no-pair-breaking
condition at a specular surface.

In the nonlinear iteration, the unknown functions are \(\Delta(\bar x)\)
and \(Q(\bar x)\).  For temporary profiles of these two functions, the
Riccati equations determine \(a_n\), \(b_n\), and hence the Green's
functions at each position.  The residuals of the gap equation and
Maxwell equation are functions of \(\bar x\):
\begin{eqnarray}
{\cal R}_{\Delta}(\bar x)
&=&
\Delta(\bar x)\ln\frac{t}{t_{c0}}
+2\pi t\sum_{n\ge0}
\left[
\frac{\Delta(\bar x)}{\Omega_n}
\right.
\nonumber\\
&&\left.
-
{\rm Re}\left\langle f_n(\bar x,\hat{\bf v})\right\rangle
\right],
\label{eq:S_gap_residual}\\
{\cal R}_{Q}(\bar x)
&=&
Q''(\bar x)+\frac{\sqrt{6}}{\pi\kappa_0^2}j_y(\bar x),
\label{eq:S_Maxwell_residual}
\end{eqnarray}
where
\begin{eqnarray}
j_y(\bar x)
&=&
-2\pi\sqrt{6}\,t\sum_{n\ge0}
\left\langle
\hat v_y\,{\rm Im}\,g_n(\bar x,\hat{\bf v})
\right\rangle .
\label{eq:S_current}
\end{eqnarray}
Here \(\langle\cdots\rangle\) denotes the average over Fermi-surface
directions.  The magnetic-field profile is obtained from $b(\bar x)=(\pi\kappa_0/\sqrt{6})Q'(\bar x)$.
Equation~(\ref{eq:S_Maxwell_residual}) follows from
\(\bar\nabla\times{\bf b}={\bf j}/\kappa_0\).

The Meissner-state solution is obtained by requiring
\begin{eqnarray}
{\cal R}_{\Delta}(\bar x)=0,
\qquad
{\cal R}_{Q}(\bar x)=0
\end{eqnarray}
at the self-consistency mesh points, together with boundary rows that
impose the conditions stated above.  In the fixed-\(Q_s\)
calculation used in this work, one imposes
\begin{eqnarray}
Q(0)=Q_s,
\qquad
Q(\bar L)=0,
\label{eq:S_fixed_Qs_BC}
\end{eqnarray}
and the surface magnetic field is then obtained from
Eq.~(\ref{eq:S_ba_Qs}).  Repeating this calculation for different values
of \(Q_s\) gives the curve \(b_a(Q_s)\).  Alternatively, in a fixed-field
formulation one imposes \(Q'(0)=(\sqrt{6}/\pi\kappa_0)b_a\) and
\(Q(\bar L)=0\).

\section{Stability against one-dimensional perturbations}
\label{sec:1D_stability}

The superheating field is determined by the loss of stability of a
Meissner-state solution.  In this section, we first consider
one-dimensional perturbations.  Here ``one-dimensional'' means that the
perturbations depend only on \(\bar x\), the distance from the surface,
and have no variation along the surface.  This is the \(\bar k=0\)
case.  Perturbations with a finite wave number along the surface are
discussed in Sec.~\ref{sec:finite_k}.

Although the Meissner-state solution itself is uniform along the
surface, the first unstable perturbation does not have to be uniform
along the surface.  A one-dimensional perturbation changes the
Meissner screening profile coherently over the whole surface.  By
contrast, a finite-\(\bar k\) perturbation allows the order parameter,
current density, and magnetic field to vary along the surface.  Such a
perturbation may be viewed as the linearized form of a
vortex-entry-like distortion near the surface.  Therefore, after
determining the \(\bar k=0\) stability limit, one must also check
whether a finite-\(\bar k\) perturbation becomes unstable at a lower
field.

\subsection{Fixed-field stability operator}
\label{sec:1D_fixed_field_operator}

For each value of \(Q_s\), Sec.~\ref{sec:nonlinear_meissner} gives a
self-consistent Meissner-state solution \(\Delta(\bar x)\), \(Q(\bar x)\),
\(j_y(\bar x)\), and \(b(\bar x)\).  We now ask whether this solution is
stable when the applied magnetic field is held fixed.  A one-dimensional
perturbation changes the order-parameter amplitude and the screening
momentum profile as
\begin{eqnarray}
\delta\Delta=\eta(\bar x),
\qquad
\delta{\bf Q}=V(\bar x)\hat{\bf y}.
\label{eq:S_1D_perturbation}
\end{eqnarray}
The corresponding magnetic-field perturbation is
\(\delta b_z(\bar x)=(\pi\kappa_0/\sqrt{6}) V'(\bar x)\).  Because the
applied field is fixed in the stability problem, the surface magnetic
field must not change.  Therefore the perturbation satisfies
\begin{eqnarray}
V'(0)=0.
\label{eq:S_1D_fixed_field_bc}
\end{eqnarray}
Fixed field does not require \(V(0)=0\).  The surface superfluid momentum
is allowed to change during a fixed-field perturbation.  Only the
surface magnetic field, which is proportional to \(Q'(0)\), must remain
fixed.  At the far end of the computational interval, the perturbation
vanishes,
\begin{eqnarray}
\eta(\bar L)=0,
\qquad
V(\bar L)=0.
\label{eq:S_1D_far_bc}
\end{eqnarray}
In the semi-infinite problem this represents
\(\eta,V\to0\) as \(\bar x\to\infty\).  No independent boundary
condition is imposed on \(\eta\) at the surface.  Its surface value is
determined by the linearized gap equation together with the reflected
quasiparticle trajectories.

To construct the stability operator, we write the Riccati amplitudes as
\(a_n=a_{0n}+\alpha_n\) and \(b_n=b_{0n}+\beta_n\), where
\(a_{0n}\) and \(b_{0n}\) are the Riccati amplitudes of the
Meissner-state solution obtained in Sec.~\ref{sec:nonlinear_meissner}.
Here \(\alpha_n\) and \(\beta_n\) are the first-order changes induced by
\(\eta\) and \(V\).  In this subsection, \(\Delta\) and \(Q\) without
the symbol \(\delta\) denote the unperturbed Meissner-state profiles.
Linearizing Eqs.~(\ref{eq:S_a_meissner}) and
(\ref{eq:S_b_meissner}) gives
\begin{eqnarray}
\pi\hat v_x\alpha_n'
&=&
(1-a_{0n}^2)\eta
-i\pi\hat v_yVa_{0n}
\nonumber\\
&&-
\left(2\Delta a_{0n}+2\Omega_n+i\pi\hat v_yQ\right)\alpha_n,
\label{eq:S_1D_alpha}\\
\pi\hat v_x\beta_n'
&=&
-(1-b_{0n}^2)\eta
+i\pi\hat v_yVb_{0n}
\nonumber\\
&&+
\left(2\Delta b_{0n}+2\Omega_n+i\pi\hat v_yQ\right)\beta_n .
\label{eq:S_1D_beta}
\end{eqnarray}
The terms proportional to \(\eta\) describe the change of the pair
potential.  The terms proportional to \(V\) describe the change of the
Doppler shift produced by the perturbation of the gauge-invariant
momentum.

Although the Meissner-state solution and the \(\bar k=0\) perturbation
depend only on \(\bar x\), the Riccati amplitudes still depend on the
Fermi-velocity direction \(\hat{\bf v}\).  For each fixed
\(\hat{\bf v}\), Eqs.~(\ref{eq:S_1D_alpha}) and
(\ref{eq:S_1D_beta}) are first-order ordinary differential equations in
\(\bar x\), with derivative terms
\(\pi\hat v_x\alpha_n'\) and \(\pi\hat v_x\beta_n'\).  We use the same
stable-integration rule as in Sec.~\ref{sec:nonlinear_meissner}:
\(\alpha_n\) is integrated in the same stable direction as \(a_n\), and
\(\beta_n\) is integrated in the same stable direction as \(b_n\).

This construction is used only to impose the linearized surface boundary
condition while keeping the Riccati integrations stable.  Since the
perturbation vanishes in the bulk, the Riccati variations whose stable
integrations start from \(\bar L\) are set to zero there and propagated
to the surface.  The remaining variations must be propagated from the
surface toward the bulk, and their surface values are fixed by specular
reflection, not chosen freely.  For the present \(\bar k=0\) perturbation,
this gives
\(\alpha_n(0;\hat v_x,\hat v_y,\hat v_z)
=\alpha_n(0;-\hat v_x,\hat v_y,\hat v_z)\) and
\(\beta_n(0;\hat v_x,\hat v_y,\hat v_z)
=\beta_n(0;-\hat v_x,\hat v_y,\hat v_z)\).  Equivalently, the
Green-function variations satisfy
\(\delta{\cal F}_n(0;\hat v_x,\hat v_y,\hat v_z)
=\delta{\cal F}_n(0;-\hat v_x,\hat v_y,\hat v_z)\), where
\(\delta{\cal F}_n\) denotes \(\delta f_n\),
\(\delta f_n^\dagger\), or \(\delta g_n\).  In this way, the linearized
Green functions for both incoming and reflected directions are obtained
without integrating any Riccati equation in its unstable direction.

The induced Green-function variations needed for the linearized
self-consistency equations are
\begin{eqnarray}
\delta f_n
&=&
\frac{2(\alpha_n-a_{0n}^2\beta_n)}
{(1+a_{0n}b_{0n})^2},
\label{eq:S_1D_df}\\
\delta g_n
&=&
-\frac{2(b_{0n}\alpha_n+a_{0n}\beta_n)}
{(1+a_{0n}b_{0n})^2}.
\label{eq:S_1D_dg}
\end{eqnarray}
Here \(\delta f_n=\delta f_n(\bar x,\hat{\bf v})\) and
\(\delta g_n=\delta g_n(\bar x,\hat{\bf v})\).  The residuals of the
linearized gap equation and Maxwell equation are
\begin{eqnarray}
{\cal R}_{\eta}(\bar x)
&=&
\eta(\bar x)\ln\frac{t}{t_{c0}}
+2\pi t\sum_{n\ge0}
\left[
\frac{\eta(\bar x)}{\Omega_n}
-
{\rm Re}\langle\delta f_n(\bar x,\hat{\bf v})\rangle
\right], \nonumber \\
\label{eq:S_1D_gap}\\
{\cal R}_{y}(\bar x)
&=&
V''(\bar x)
+\frac{\sqrt{6}}{\pi\kappa_0^2}\delta j_y(\bar x).
\label{eq:S_1D_Maxwell}
\end{eqnarray}
The current response appearing in the Maxwell residual is
\begin{eqnarray}
\delta j_y(\bar x)
&=&
-2\pi\sqrt{6}\,t\sum_{n\ge0}
\left\langle
\hat v_y\,{\rm Im}\,\delta g_n(\bar x,\hat{\bf v})
\right\rangle .
\label{eq:S_1D_dj}
\end{eqnarray}
Equations~(\ref{eq:S_1D_alpha})--(\ref{eq:S_1D_dj}), together with
the boundary conditions in Eqs.~(\ref{eq:S_1D_fixed_field_bc}) and
(\ref{eq:S_1D_far_bc}), define a homogeneous fixed-field stability
problem for \(\eta(\bar x)\) and \(V(\bar x)\).

After discretization, \(\eta(\bar x)\) and \(V(\bar x)\) are not two
numbers, but two sets of values on the \(\bar x\)-mesh.  If the mesh
points are denoted by \(\bar x_i\), \(i=1,\ldots,N\), we define
\begin{eqnarray}
{\bf u}_{\rm 1D}
&=&
\left(
\eta_1,\ldots,\eta_N,
V_1,\ldots,V_N
\right)^T .
\label{eq:S_u1D_discrete}
\end{eqnarray}
Here \(\eta_i :=\eta(\bar x_i)\) and \(V_i :=V(\bar x_i)\).
The corresponding residual vector is
\begin{eqnarray}
{\bf r}_{\rm 1D}
=
\left(
{\cal R}_{\eta,1},\ldots,{\cal R}_{\eta,N},
\right.
\left.
{\cal R}_{y,1},\ldots,{\cal R}_{y,N}
\right)^T .
\label{eq:S_r1D_discrete}
\end{eqnarray}
The vector \({\bf r}_{\rm 1D}\) contains both equation residuals and
boundary-condition residuals.  At interior mesh points, its components
are the discretized linearized gap and Maxwell residuals,
\({\cal R}_{\eta,i}\) and \({\cal R}_{y,i}\).  Rows associated with
imposed boundary conditions are used to enforce conditions such as
\(V'(0)=0\), \(\eta(\bar L)=0\), and \(V(\bar L)=0\).  Thus
\({\bf r}_{\rm 1D}=0\) means that both the linearized self-consistency
equations and the boundary conditions are satisfied.

Linearizing the self-consistency equations defines a linear map from
\({\bf u}_{\rm 1D}\) to \({\bf r}_{\rm 1D}\):
\begin{eqnarray}
{\bf r}_{\rm 1D}={\cal M}_{\rm 1D}{\bf u}_{\rm 1D}.
\label{eq:S_M1D_def}
\end{eqnarray}
Thus \({\cal M}_{\rm 1D}\) is a \(2N\times2N\) stability operator, not a
\(2\times2\) matrix.  
The calculation does not assume a particular
shape for \(\eta(\bar x)\) or \(V(\bar x)\).  The instability profile is
found from this discretized linear problem.

The loss of stability means that a nonzero infinitesimal deformation of
the Meissner-state solution can satisfy all linearized equations and
boundary conditions at fixed applied field.  In the discretized problem,
such a deformation is represented by a nonzero vector
\({\bf u}_{\rm 1D}\).  The vector
\({\cal M}_{\rm 1D}{\bf u}_{\rm 1D}\) is the residual produced by this
deformation.  If the Meissner-state solution is stable, the only
deformation with zero residual is the trivial one,
\({\bf u}_{\rm 1D}=0\).  At the stability limit, a nontrivial
deformation with \({\bf u}_{\rm 1D}\ne0\) appears and satisfies
\({\cal M}_{\rm 1D}{\bf u}_{\rm 1D}=0\). 
We refer to this perturbation as the zero mode.

To detect this condition numerically, we ask how small the residual can
become among normalized perturbations.  We define
\begin{eqnarray}
s_{\rm min}({\cal M}_{\rm 1D})
=
\min_{\|{\bf u}_{\rm 1D}\|=1}
\|{\cal M}_{\rm 1D}{\bf u}_{\rm 1D}\| .
\label{eq:S_smin_definition}
\end{eqnarray}
This quantity is the smallest singular value of
\({\cal M}_{\rm 1D}\).  If every normalized perturbation gives a finite
residual, then \(s_{\rm min}({\cal M}_{\rm 1D})>0\).  If a nonzero
perturbation satisfies the linearized gap equation, Maxwell equation,
and boundary conditions, then the residual can vanish and
\(s_{\rm min}({\cal M}_{\rm 1D})=0\).  Therefore the condition
\begin{eqnarray}
s_{\rm min}({\cal M}_{\rm 1D})\to0
\label{eq:S_1D_smin}
\end{eqnarray}
is the numerical form of the one-dimensional zero-mode condition.  The
corresponding singular vector gives the spatial profile of the
incipient instability.

Finally, we note that \({\cal M}_{\rm 1D}\) is not just a local
finite-difference matrix.  The local Maxwell term, such as
\(V''(\bar x)\), connects only nearby mesh points after discretization.
In contrast, the clean Eilenberger response connects distant mesh points.
This is because the effect of a perturbation in \(\eta\) or \(V\) is
propagated through the linearized Riccati equations before the gap and
current residuals are evaluated.  As a result, the residual at a mesh
point \(\bar x_i\) can depend on \(\eta_j\) and \(V_j\) at distant mesh
points.  In matrix form, this appears as broad off-diagonal matrix
elements of \({\cal M}_{\rm 1D}\).

\subsection{Relation to the \texorpdfstring{\(Q_s\)}{Qs} scan}
\label{sec:1D_Qs_scan}

We now explain how this fixed-field stability problem is related to the
\(Q_s\) scan described in Sec.~\ref{sec:nonlinear_meissner}.  In the
fixed-\(Q_s\) calculation, one imposes \(Q(0)=Q_s\), solves the nonlinear
self-consistency problem, and then obtains the surface field from
\(b_a(Q_s)=(\pi\kappa_0/\sqrt{6}) Q'(0)\).  Thus the numerical
calculation gives a curve \(b_a(Q_s)\). 

Consider two nearby self-consistent Meissner-state solutions with
surface momenta \(Q_s\) and \(Q_s+\delta Q_s\).  Their difference gives
\begin{eqnarray}
\eta_*(\bar x)
&:=&
\frac{\partial \Delta(\bar x;Q_s)}{\partial Q_s}\delta Q_s,
\label{eq:S_tangent_eta}\\
V_*(\bar x)
&:=&
\frac{\partial Q(\bar x;Q_s)}{\partial Q_s}\delta Q_s .
\label{eq:S_tangent_V}
\end{eqnarray}
Because both solutions satisfy the same self-consistency equations, this
difference satisfies the corresponding linearized equations.  Its
boundary values are
\begin{eqnarray}
V_*(0)=\delta Q_s,
\qquad
V_*(\bar L)=0,
\qquad
\eta_*(\bar L)=0.
\label{eq:S_tangent_bc}
\end{eqnarray}
The only remaining condition for this difference to be an allowed
fixed-field perturbation is \(V_*'(0)=0\).  Since
\(b_a(Q_s)=(\pi\kappa_0/\sqrt{6})Q'(0;Q_s)\), the change of the surface
field along the curve is
\begin{eqnarray}
\delta b_a
=
\frac{\pi\kappa_0}{\sqrt{6}}V_*'(0)
=
\frac{d b_a}{dQ_s}\delta Q_s .
\label{eq:S_delta_ba}
\end{eqnarray}
Therefore, away from an extremum of \(b_a(Q_s)\), the difference between
nearby fixed-\(Q_s\) solutions changes the applied field and is not an
allowed fixed-field perturbation.  At an extremum of \(b_a(Q_s)\),
however, \(d b_a/dQ_s=0\), and Eq.~(\ref{eq:S_delta_ba}) gives
\(V_*'(0)=0\).  At that point the same difference between nearby
solutions satisfies all fixed-field perturbation boundary conditions.  It
is nonzero, because \(V_*(0)=\delta Q_s\), and it solves the homogeneous
linearized equations.  Hence it is a zero mode of the fixed-field
stability operator:
\begin{eqnarray}
&&{\cal M}_{\rm 1D}{\bf u}_{*,{\rm 1D}}=0, \\
&&{\bf u}_{*,{\rm 1D}}
=
\left(
\eta_{*,1},\ldots,\eta_{*,N},
V_{*,1},\ldots,V_{*,N}
\right)^T .
\label{eq:S_1D_zero_mode}
\end{eqnarray}

This argument also explains why the first maximum gives the
one-dimensional fixed-field stability limit.  In the present calculation,
the self-consistent Meissner-state solutions at fixed \(Q_s\) form a
smooth locally unique curve connected to the weak-field state.  Within
this locally unique set of solutions, the infinitesimal displacement
that keeps the solution self-consistent is the tangent to this curve.
This tangent becomes an allowed fixed-field perturbation only when
\(d b_a/dQ_s=0\).  Therefore the first maximum of \(b_a(Q_s)\), reached
as the field is increased from zero, gives the one-dimensional
fixed-field stability limit.  
Therefore the one-dimensional stability limit is
\begin{eqnarray}
B_{\rm sh}^{\rm 1D}= \max_{Q_s} B_s(Q_s).
\label{eq:S_Hsh1D}
\end{eqnarray}


\section{Finite-wave-number stability operator}
\label{sec:finite_k}

The previous section treated perturbations that are uniform along the
surface.  We now consider perturbations with a finite wave number
\(\bar k>0\) along the surface.  This calculation checks whether a
surface-modulated perturbation becomes unstable before the
\(\bar k=0\) instability discussed in Sec.~\ref{sec:1D_stability}.

We use \(\bar y=y/\xi_0\) and write the perturbation as
\begin{eqnarray}
\delta\Delta(\bar x,\bar y)
&=&
\eta(\bar x)\cos \bar k\bar y,
\label{eq:S_dDelta}\\
\delta{\bf Q}(\bar x,\bar y)
&=&
W(\bar x)\sin\bar k\bar y\,\hat{\bf x}
+
V(\bar x)\cos\bar k\bar y\,\hat{\bf y}.
\label{eq:S_dQ}
\end{eqnarray}
The relative sine and cosine phases are chosen so that the magnetic
perturbation has a single cosine harmonic.  Taking the rotation of
\(\delta{\bf Q}\) gives
\begin{eqnarray}
\delta b_z(\bar x,\bar y)
=
\frac{\pi\kappa_0}{\sqrt{6}}
\left[V'(\bar x)-\bar kW(\bar x)\right]
\cos\bar k\bar y .
\label{eq:S_dbz}
\end{eqnarray}
It is useful to define \(h(\bar x) :=V'(\bar x)-\bar kW(\bar x)\), so
that \(h\) is proportional to the magnetic-field perturbation.  Since
the applied field is fixed in the stability problem, the surface
magnetic field must not change.  Therefore
\begin{eqnarray}
h(0)=V'(0)-\bar kW(0)=0.
\label{eq:S_fixed_field_2D}
\end{eqnarray}
This fixed-field condition does not require \(W(0)=0\) or \(V(0)=0\).
The surface values of \(W\) and \(V\) are determined by the stability
problem itself.  At the far end of the computational interval, the
perturbation is required to vanish:
\begin{eqnarray}
\eta(\bar L)=0,
\qquad
W(\bar L)=0,
\qquad
V(\bar L)=0.
\label{eq:S_2D_far_bc}
\end{eqnarray}

The finite-\(\bar k\) problem differs from the \(\bar k=0\) problem
because the perturbation has a prescribed dependence on the surface
coordinate \(\bar y\).  In the Eilenberger equation, the derivative along
the Fermi-velocity direction contains
\(\hat{\bf v}\cdot\bar\nabla
=\hat v_x\partial_{\bar x}+\hat v_y\partial_{\bar y}\).  Here
\(\hat v_x\) and \(\hat v_y\) are scalar components of the unit
Fermi-velocity direction \(\hat{\bf v}\).  After substituting the
harmonic forms proportional to \(\cos\bar k\bar y\) and
\(\sin\bar k\bar y\), the unknown amplitudes depend only on \(\bar x\).
Thus \(\partial_{\bar x}\) becomes an ordinary derivative of these
amplitudes, denoted by a prime, while \(\partial_{\bar y}\) is evaluated
analytically: it converts a cosine harmonic into a sine harmonic, and a
sine harmonic into a cosine harmonic.  As a result, the Riccati
perturbations must contain both harmonics.
We write
\begin{eqnarray}
a_n
&=&
a_{0n}+\alpha_{cn}\cos\bar k\bar y+\alpha_{sn}\sin\bar k\bar y,
\label{eq:S_a_expand}\\
b_n
&=&
b_{0n}+\beta_{cn}\cos\bar k\bar y+\beta_{sn}\sin\bar k\bar y .
\label{eq:S_b_expand}
\end{eqnarray}
Here \(a_{0n}\) and \(b_{0n}\) are the Riccati amplitudes of the
Meissner-state solution obtained in Sec.~\ref{sec:nonlinear_meissner}.
The amplitudes \(\alpha_{cn}\), \(\alpha_{sn}\), \(\beta_{cn}\), and
\(\beta_{sn}\) are functions of \(\bar x\) and of the Fermi-velocity
direction, although these arguments are not written explicitly.  The
subscript \(c\) denotes the component multiplying
\(\cos\bar k\bar y\), and the subscript \(s\) denotes the component
multiplying \(\sin\bar k\bar y\).  In this section, \(\Delta\) and
\(Q\) without the symbol \(\delta\) denote the unperturbed
Meissner-state profiles.

Linearizing the Riccati equations and equating the cosine and sine
components gives
\begin{eqnarray}
\pi(\hat v_x\alpha_{cn}'+\bar k\hat v_y\alpha_{sn})
&=&
-\left(2\Delta a_{0n}+2\Omega_n+i\pi\hat v_yQ\right)\alpha_{cn}
\nonumber\\
&&-i\pi\hat v_yVa_{0n}
+(1-a_{0n}^2)\eta,
\label{eq:S_alpha_c}\\
\pi(\hat v_x\beta_{cn}'+\bar k\hat v_y\beta_{sn})
&=&
\left(2\Delta b_{0n}+2\Omega_n+i\pi\hat v_yQ\right)\beta_{cn}
\nonumber\\
&&+i\pi\hat v_yVb_{0n}
-(1-b_{0n}^2)\eta,
\label{eq:S_beta_c}\\
\pi(\hat v_x\alpha_{sn}'-\bar k\hat v_y\alpha_{cn})
&=&
-\left(2\Delta a_{0n}+2\Omega_n+i\pi\hat v_yQ\right)\alpha_{sn}
\nonumber\\
&&-i\pi\hat v_xWa_{0n},
\label{eq:S_alpha_s}\\
\pi(\hat v_x\beta_{sn}'-\bar k\hat v_y\beta_{cn})
&=&
\left(2\Delta b_{0n}+2\Omega_n+i\pi\hat v_yQ\right)\beta_{sn}
\nonumber\\
&&+i\pi\hat v_xWb_{0n}.
\label{eq:S_beta_s}
\end{eqnarray}
The terms proportional to \(\bar k\hat v_y\) come from the
\(\bar y\)-derivative and describe the coupling between the cosine and
sine components.  The terms proportional to \(V\) and \(W\) come from
the perturbation of \(\hat{\bf v}\cdot{\bf Q}\).
For each fixed Fermi-velocity direction \(\hat{\bf v}\),
Eqs.~(\ref{eq:S_alpha_c}) and (\ref{eq:S_alpha_s}) are solved as a
coupled pair of first-order equations for \(\alpha_{cn}\) and
\(\alpha_{sn}\).  They are integrated in the same stable direction as
\(a_n\).  Similarly, Eqs.~(\ref{eq:S_beta_c}) and
(\ref{eq:S_beta_s}) are solved as a coupled pair for \(\beta_{cn}\) and
\(\beta_{sn}\), and are integrated in the same stable direction as
\(b_n\).  Thus, for positive Matsubara frequencies, the \(\alpha\)
components are propagated from the surface to the bulk for
\(\hat v_x>0\), and from the bulk to the surface for \(\hat v_x<0\).
The \(\beta\) components are propagated in the opposite directions.
As in Sec.~\ref{sec:nonlinear_meissner}, the components whose stable
integration starts from the bulk use zero perturbation in the bulk
asymptotic state, and the surface values of the partner components are
obtained from specular reflection.

At the surface, the linearized Green functions satisfy
\begin{eqnarray}
\delta{\cal F}_n(0,\bar y;\hat v_x,\hat v_y,\hat v_z)
=
\delta{\cal F}_n(0,\bar y;-\hat v_x,\hat v_y,\hat v_z),
\label{eq:S_2D_specular_Green}
\end{eqnarray}
where \(\delta{\cal F}_n\) denotes \(\delta f_n\),
\(\delta f_n^\dagger\), or \(\delta g_n\).  
Since the reflected direction has the same \(\bar y\), the cosine and
sine amplitudes obey the same reflection rule separately.
Equivalently,
\(\alpha_{cn}\), \(\alpha_{sn}\), \(\beta_{cn}\), and \(\beta_{sn}\)
at \((0;\hat v_x,\hat v_y,\hat v_z)\) are equal to the corresponding
amplitudes at \((0;-\hat v_x,\hat v_y,\hat v_z)\).

The induced Green-function harmonics are
\begin{eqnarray}
\delta f_{cn}
&=&
\frac{2(\alpha_{cn}-a_{0n}^2\beta_{cn})}
{(1+a_{0n}b_{0n})^2},
\label{eq:S_dfc}\\
\delta f_{sn}
&=&
\frac{2(\alpha_{sn}-a_{0n}^2\beta_{sn})}
{(1+a_{0n}b_{0n})^2},
\label{eq:S_dfs}\\
\delta g_{cn}
&=&
-\frac{2(b_{0n}\alpha_{cn}+a_{0n}\beta_{cn})}
{(1+a_{0n}b_{0n})^2},
\label{eq:S_dgc}\\
\delta g_{sn}
&=&
-\frac{2(b_{0n}\alpha_{sn}+a_{0n}\beta_{sn})}
{(1+a_{0n}b_{0n})^2}.
\label{eq:S_dgs}
\end{eqnarray}
The gap-amplitude perturbation is chosen to be proportional to \(\cos\bar k\bar y\).  Therefore the real cosine component of the
complex gap equation gives the amplitude self-consistency condition. We choose the phase gauge so that the order-parameter perturbation has no sine component.  The imaginary sine component is not an additional gap-amplitude equation; it is used below as a surface phase
self-consistency condition.
The current response has the two harmonics needed by Maxwell's equations:
\begin{eqnarray}
\delta j_x^{(s)}(\bar x)
&=&
-2\pi\sqrt{6}\,t\sum_{n\ge0}
\left\langle \hat v_x\,{\rm Im}\,\delta g_{sn}(\bar x,\hat{\bf v})\right\rangle,
\label{eq:S_djx}\\
\delta j_y^{(c)}(\bar x)
&=&
-2\pi\sqrt{6}\,t\sum_{n\ge0}
\left\langle \hat v_y\,{\rm Im}\,\delta g_{cn}(\bar x,\hat{\bf v})\right\rangle .
\label{eq:S_djy}
\end{eqnarray}
The superscripts \((s)\) and \((c)\) indicate the sine and cosine
harmonics, respectively.

Using \(h=V'-\bar kW\), the linearized Maxwell equations are obtained
from \(\bar\nabla\times\delta{\bf b}=\delta{\bf j}/\kappa_0\).  The
\(x\) component is the equation for the sine harmonic of the current,
and the \(y\) component is the equation for the cosine harmonic.  Together
with the gap equation, the residuals are
\begin{eqnarray}
{\cal R}_{\eta}(\bar x)
&=&
\eta(\bar x)\ln\frac{t}{t_{c0}} \nonumber \\
&&+
2\pi t\sum_{n\ge0}
\left[
\frac{\eta(\bar x)}{\Omega_n}
-
{\rm Re}\langle\delta f_{cn}(\bar x,\hat{\bf v})\rangle
\right],
\label{eq:S_lin_gap}\\
{\cal R}_{x}(\bar x)
&=&
-\bar k\left[V'(\bar x)-\bar kW(\bar x)\right]
-
\frac{\sqrt{6}}{\pi\kappa_0^2}\delta j_x^{(s)}(\bar x),
\label{eq:S_lin_Maxwell_x}\\
{\cal R}_{y}(\bar x)
&=&
-\left[V'(\bar x)-\bar kW(\bar x)\right]'
-
\frac{\sqrt{6}}{\pi\kappa_0^2}\delta j_y^{(c)}(\bar x).
\label{eq:S_lin_Maxwell_y}
\end{eqnarray}

For finite \(\bar k>0\), specular reflection gives
\(\delta j_x^{(s)}(0)=0\).  Therefore
Eq.~(\ref{eq:S_lin_Maxwell_x}) gives
\({\cal R}_x(0)=-\bar k h(0)\), and the surface condition
\({\cal R}_x(0)=0\) is not independent of the fixed-field condition
\(h(0)=0\).  In the present calculation, the surface
\({\cal R}_x\) row is replaced by the imaginary sine component of the
microscopic gap self-consistency equation,
\begin{eqnarray}
{\cal R}_{\rm ph}^{(s)}(0)
=
-2\pi t\sum_{n\ge0}
{\rm Im}\left\langle
\delta f_{sn}(0,\hat{\bf v})
\right\rangle
=0 .
\label{eq:S_phase_surface}
\end{eqnarray}
In the interior, this phase condition is not added as a separate
equation because it is related to current conservation through the
linearized Maxwell equations.  The \(\bar k=0\) problem is treated
separately in Sec.~\ref{sec:1D_stability}.

After discretization, the unknown functions \(\eta(\bar x)\),
\(W(\bar x)\), and \(V(\bar x)\) become sets of values on the
\(\bar x\)-mesh.  With mesh points \(\bar x_i\), \(i=1,\ldots,N\), we
define
\begin{eqnarray}
{\bf u}_{\bar k}
&=&
\left(
\eta_1,\ldots,\eta_N,
W_1,\ldots,W_N,
V_1,\ldots,V_N
\right)^T.
\label{eq:S_uk_discrete}
\end{eqnarray}
The corresponding residual vector is
\begin{eqnarray}
{\bf r}_{\bar k}
&=&
\left(
{\cal R}_{\eta,1},\ldots,{\cal R}_{\eta,N},
\right.
\nonumber\\
&&\left.
{\cal R}_{x,1},\ldots,{\cal R}_{x,N},
{\cal R}_{y,1},\ldots,{\cal R}_{y,N}
\right)^T.
\label{eq:S_rk_discrete}
\end{eqnarray}
In the block notation of Eq.~(\ref{eq:S_rk_discrete}), the surface
entry of the \({\cal R}_x\) block is replaced by
\({\cal R}_{\rm ph}^{(s)}(0)\), while the surface entry of the
\({\cal R}_y\) block is replaced by \(h(0)\).
The residual vector contains both equation residuals and
boundary-condition residuals.  At interior mesh points, its components
are the discretized forms of
\({\cal R}_{\eta}(\bar x)\), \({\cal R}_{x}(\bar x)\), and
\({\cal R}_{y}(\bar x)\).  
The far-boundary rows impose
\(\eta(\bar L)=W(\bar L)=V(\bar L)=0\).  Thus
\({\bf r}_{\bar k}=0\) means that both the linearized equations and the
boundary conditions are satisfied.

Linearizing the self-consistency equations
therefore defines
\begin{eqnarray}
{\bf r}_{\bar k}={\cal M}_{\bar k}{\bf u}_{\bar k}.
\label{eq:S_Mk_def}
\end{eqnarray}
Thus \({\cal M}_{\bar k}\) is a \(3N\times3N\) stability operator, not a
\(3\times3\) matrix.  No particular \(\bar x\)-profile is assumed for
\(\eta\), \(W\), or \(V\).  As in the \(\bar k=0\) case, the clean
quasiparticle response makes \({\cal M}_{\bar k}\) nonlocal.  In the
discretized problem, this means that the residual at one mesh point can
depend on perturbation amplitudes at distant mesh points, so
\({\cal M}_{\bar k}\) contains broad off-diagonal matrix elements.

For a fixed Meissner-state solution and a fixed \(\bar k\), an
instability is reached when the linearized problem has a nonzero
solution with zero residual.  Numerically, this is detected by the same
minimum-residual criterion as in Sec.~\ref{sec:1D_fixed_field_operator}.
Equivalently, we monitor the smallest singular value,
\begin{eqnarray}
s_{\rm min}({\cal M}_{\bar k})\to0.
\label{eq:S_smin_k}
\end{eqnarray}
The \(\bar k=0\) point is the one-dimensional stability problem of
Sec.~\ref{sec:1D_stability}.  The full superheating field is therefore
\begin{eqnarray}
B_{\rm sh}=\min_{\bar k\ge0}B_{\rm inst}(\bar k),
\label{eq:S_full_Bsh_min_k}
\end{eqnarray}
where \(B_{\rm inst}(\bar k)\) is the first field at which
\({\cal M}_{\bar k}\) has a null vector.

In the calculations reported in the Letter, the finite-\(\bar k\)
analysis is used as a check on the one-dimensional candidate obtained
from Sec.~\ref{sec:1D_stability}.  For the low-\(\kappa\) parameter
range studied here, no finite-\(\bar k\) perturbation becomes unstable
below \(B_{\rm sh}^{\rm 1D}\).  Hence we find that, in the parameter range studied here, the
superheating field is given by the one-dimensional value,
\(B_{\rm sh}=B_{\rm sh}^{\rm 1D}\).

\end{document}